\documentclass{article}

\usepackage[english]{babel}

\usepackage[letterpaper,top=2cm,bottom=2cm,left=3cm,right=3cm,marginparwidth=1.75cm]{geometry}

\usepackage{amsmath}
\usepackage{graphicx}
\usepackage[colorlinks=true, allcolors=blue]{hyperref}
\usepackage{natbib}
\usepackage{multirow}

\title{CodeShell TECHNICAL REPORT}
\author{
	Rui Xie, Zhengran Zeng, Zhuohao Yu, Chang Gao, Shikun Zhang, Wei Ye \\
 National Engineering Research Center for Software Engineering, Peking University, China \\
 \{ruixie, wye\}@pku.edu.cn \\
 https://github.com/WisdomShell/codeshell
}
\date{}

\begin{document}
\maketitle

\begin{abstract}

Code large language models mark a pivotal breakthrough in artificial intelligence. They are specifically crafted to understand and generate programming languages, significantly boosting the efficiency of coding development workflows. In this technical report, we present CodeShell-Base, a seven billion-parameter foundation model with 8K context length, showcasing exceptional proficiency in code comprehension. By incorporating Grouped-Query Attention and Rotary Positional Embedding into GPT-2, CodeShell-Base integrates the structural merits of StarCoder and CodeLlama and forms its unique architectural design. We then carefully built a comprehensive data pre-processing process, including similar data deduplication, perplexity-based data filtering, and model-based data filtering. Through this process, We have curated 100 billion high-quality pre-training data from GitHub. Benefiting from the high-quality data, CodeShell-Base outperforms CodeLlama in Humaneval after training on just 500 billion tokens (5 epochs). We have conducted extensive experiments across multiple language datasets, including Python, Java, and C++, and the results indicate that our model possesses robust foundational capabilities in code comprehension and generation.

\end{abstract}

\section{Introduction}

Code Large Language Models (CodeLLM) such as CodeGen~\citep{codegen}, CodeLlama~\citep{codellama}, and StarCoder~\citep{starcoder} have revolutionized software development by automating tasks, reducing errors, and improving efficiency~\citep{gpt4report}. Leveraging deep learning and vast code datasets~\citep{codegen,thestack}, these models enhance developer productivity and make software development more accessible to a broader audience.

Current CodeLLMs can broadly be divided into three main categories: pre-training from scratch~\citep{starcoder}, pre-training from an existing LLM~\citep{codex,codellama}, and Instruct Tuning~\citep{wizardcoder}. Models that are pre-trained from scratch require a substantial volume of data and considerable time~\citep{starcoder,llama2}. On the other hand, pre-training from an existing LLM leverages a pre-existing model as its foundation, which brings the benefits of shortened training times and improved efficiency with less data~\citep{codex,codellama}. Instruct Tuning, meanwhile, involves the fine-tuning of an already large model using instructive data, with the goal of significantly enhancing the model's performance~\citep{codellama,wizardcoder}. However, a significant challenge in this area is that existing large models are trained on vast datasets of code without meticulous data governance, which could lead to the production of low-quality code. Although there have been some code selection stratigies~\citep{phi1} from the training model perspective, the risk of generating low-quality code remains a concern.

In this technical report, we introduce a large code model named CodeShell. CodeShell incorporated  Rotary Positional Embedding (ROPE)~\citep{rope} and  grouped-query attention~\citep{gqa} into GPT-2~\citep{gpt2}, building an efficient and context-expansion-friendly architecture. We then developed a pipeline for high-quality code selection, resulting in the acquisition of 100 billion tokens of high-quality code. Building on this foundation of 100 billion tokens, CodeShell was trained over five epochs. Our experiments demonstrate that training solely on 100 billion unique tokens of code enables CodeShell to achieve performance that is on par with, if not superior to, existing large models. For context, StarCoder~\citep{starcoder}  and CodeLlama~\citep{codellama} were both trained on 250 billion unique tokens. As is well known, high-quality code is limited within the existing open-source repositories, making the selection of high-quality code critically important for developing a high-quality code model. Our main contributions are:

\begin{itemize}

\item We released CodeShell-7B, a new large code foundation model pre-trained from scratch featuring a novel and unique architecture design.Through tests on a variety of public code-related benchmarks, it has demonstrated competitive performance across multiple programming languages.

\item In order to reduce the training costs of large code models, we have constructed an efficient data preprocessing pipeline capable of identifying high-quality code snippets from massive code corpora. Experimental results indicate that our model, trained on only 500B tokens, surpasses the performance of StarCoder trained on 1 trillion tokens. 

\item To address more complex and extensive coding tasks~\citep{codereval,defects4j}, we have increased the model's context length to 8K, enhancing its capability to process longer code segments. Empirical evidence indicates that the extension of pre-training phases with augmented lengths of code significantly improves the model's proficiency in managing extended code sequences, without detrimentally affecting its efficacy in handling shorter code snippets.

\end{itemize}

\section{CodeShell}

\subsection{Data}

To further elaborate on the details provided in the context of constructing the CodeShell training dataset and its subsequent filtering and enhancement process, let's dive into each step with more specific information:

\textbf{Data Collection:} Our primary source, GitHub~\citep{gharchive}, was meticulously scanned to amass a comprehensive collection of repositories. This included a direct crawl of 15TB of GitHub repositories to ensure a broad and diverse dataset.Inclusion of the Stack~\citep{thestack} and StarCoder datasets provided a rich variety of code examples and programming discussions, enriching the diversity of the training material.

\textbf{Language Filtering:} The decision to exclude languages with dataset volumes smaller than 100 MB was strategic, focusing on languages with significant enough usage and examples to contribute meaningful learning patterns. The incorporation of 7 code-related corpus, such as markdown for documentation, git-commits for development practices, and GitHub-issues for problem-solving discussions, aimed to broaden the model's understanding of programming ecosystems.

\textbf{Initial Filtering Rules:} The exclusion of code with excessive line lengths and low alphabetic character content was designed to eliminate atypical, potentially non-representative data, focusing the training on more standard and readable code examples.

\textbf{De-duplication:} The application of MD5 hashing allowed for the efficient identification and removal of exact duplicate texts, while MinHash~\citep{minihash} techniques were employed to detect and filter out highly similar, but not identical, content. This step was crucial for enhancing dataset diversity and quality.

\textbf{Perplexity-Based Filtering:} Perplexity scores~\citep{pplf} were utilized as a measure of the predictability and quality of code texts. By filtering out code with high perplexity scores, we aimed to exclude low-quality or confusing code examples, refining the dataset quality further.


\textbf{Rule-Based Filtering:} Our rule-based analysis and filtering system enables the effective identification and selection of high-quality code examples through a customized screening process. Specifically, we begin by defining and calculating a series of metrics for code files, including the number of lines of code, the presence and extent of comments, and the average number of characters per line. We then establish a threshold to isolate exemplary code instances that meet our criteria.
Furthermore, the system introduces a preferential bias toward code that utilizes well-known third-party libraries, an indicative marker of robust development practices. Additionally, we emphasize the selection of code that exhibits a suitable level of logical complexity. This is determined through the in-depth analysis of the abstract syntax tree and control flow within the codes, aiming to ensure that the examples we choose are not only functional but also sophisticated.
Another crucial aspect is the adherence of the code to established coding standards. Maintaining compliance with these standards is considered a reflection of best practices within the industry.

\begin{figure}[hbt]
\centering
\includegraphics[width=0.8\linewidth]{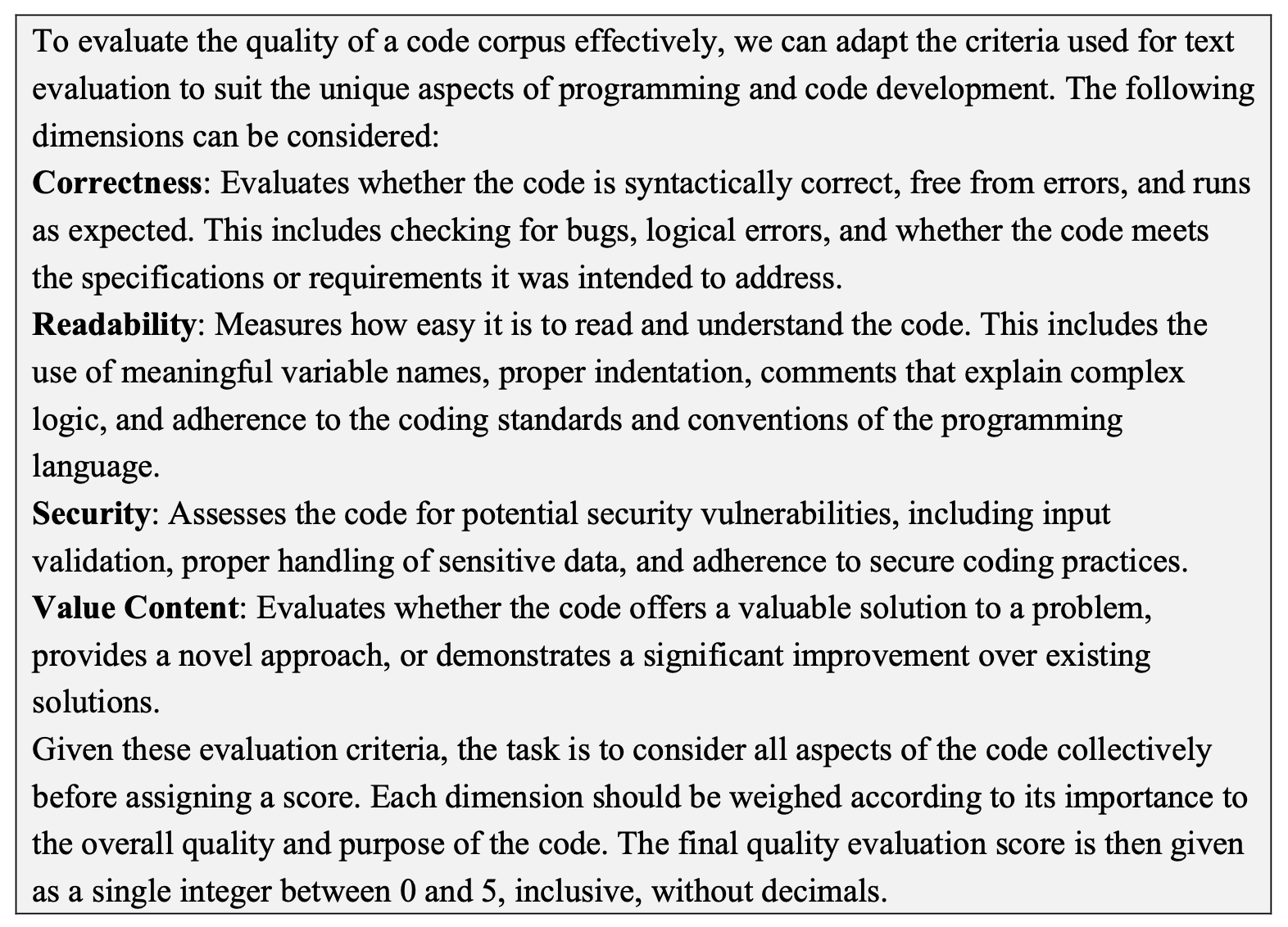}
\caption{Prompt of code quality annotation.}
\label{fig:prompt_code_anno}
\end{figure}

\textbf{Learning-Based Filtering:} The deep learning-based filter was an innovative step beyond traditional educational content filters. By training a BERT-based model on a dataset scored for code quality, we could automate the evaluation of code correctness, readability, security, and whether the code offers a valuable solution to a problem. 
The code quality dataset is generated by GPT4, Figure \ref{fig:prompt_code_anno} shows an example prompt for acode quality annotation.
This approach allowed for a nuanced and sophisticated selection of high-quality code, substantially improving the dataset's overall quality.

This detailed process, from initial collection through sophisticated filtering techniques, underscores our commitment to crafting a high-quality, diverse training dataset for CodeShell. By employing this hybrid method, we are able to more accurately identify and extract instances of high-quality code, thereby providing a superior data foundation for training advanced code models like CodeShell. Moreover, the application of this method extends beyond enhancing model performance. The final data distribution is illustrated in the Table \ref{tab:data_overview_1}. Due to page limits, we have only showcased the top 50 languages by the number of tokens.

\begin{table}[!ht]
	\centering
    \caption{
    Overview of the training data for CodeShell. For the selected programming languages, we show the number of files and data volume before and after data cleaning.
    }
	\begin{tabular}{lcccccc}
    \hline
    \multirow{2}*{Language}&\multicolumn{3}{c}{\textbf{Before}}&\multicolumn{3}{c}{\textbf{After}} \\
    \cline{2-7}
    & \small{Size (GB)}  & \small{Files (k)} & \small{Prop. (\%)} & \small{Tokens (B)}  & \small{Files (k)} & \small{Prop. (\%)} \\
    \hline
markdown & 109.01 & 22017.40 & 8.64\% & 9.80 & 1980.15 & 20.18\% \\
python & 69.56 & 12829.11 & 5.51\% & 6.70 & 1235.80 & 13.79\% \\
java & 112.93 & 23530.24 & 8.95\% & 5.82 & 1212.83 & 11.98\% \\
javascript & 141.63 & 34203.14 & 11.23\% & 5.73 & 1382.77 & 11.79\% \\
php & 130.06 & 28880.88 & 10.31\% & 5.25 & 1164.92 & 10.80\% \\
cpp & 127.14 & 15583.21 & 10.08\% & 4.21 & 515.61 & 8.66\% \\
c & 51.52 & 7127.18 & 4.08\% & 2.27 & 313.44 & 4.66\% \\
c-sharp & 96.97 & 20324.36 & 7.69\% & 1.66 & 347.48 & 3.41\% \\
go & 35.90 & 6686.82 & 2.85\% & 0.99 & 184.60 & 2.04\% \\
rust & 15.12 & 2082.10 & 1.20\% & 0.92 & 127.31 & 1.90\% \\
ruby & 7.29 & 2557.87 & 0.58\% & 0.60 & 209.89 & 1.23\% \\
kotlin & 6.47 & 2054.30 & 0.51\% & 0.51 & 160.45 & 1.04\% \\
swift & 3.04 & 560.94 & 0.24\% & 0.46 & 85.10 & 0.95\% \\
typescript & 9.44 & 2864.80 & 0.75\% & 0.44 & 133.40 & 0.90\% \\
scala & 8.93 & 2264.89 & 0.71\% & 0.41 & 104.94 & 0.85\% \\
lua & 5.24 & 811.57 & 0.42\% & 0.39 & 59.82 & 0.79\% \\
julia & 2.94 & 575.04 & 0.23\% & 0.35 & 67.59 & 0.71\% \\
github-issues-filtered-structured & 59.81 & 26747.93 & 4.74\% & 0.27 & 119.20 & 0.55\% \\
html & 55.38 & 5914.34 & 4.39\% & 0.25 & 26.36 & 0.51\% \\
git-commits-cleaned & 49.81 & 7138.09 & 3.95\% & 0.22 & 31.81 & 0.46\% \\
shell & 6.59 & 2715.93 & 0.52\% & 0.20 & 84.16 & 0.42\% \\
perl & 4.17 & 595.53 & 0.33\% & 0.19 & 27.87 & 0.40\% \\
css & 30.09 & 5292.28 & 2.38\% & 0.13 & 23.58 & 0.28\% \\
assembly & 4.21 & 553.85 & 0.33\% & 0.13 & 17.02 & 0.27\% \\
powershell & 2.84 & 586.14 & 0.22\% & 0.10 & 19.76 & 0.20\% \\
json & 15.91 & 8939.52 & 1.26\% & 0.07 & 39.84 & 0.15\% \\
r & 0.59 & 77.32 & 0.05\% & 0.06 & 8.38 & 0.13\% \\
sql & 12.69 & 784.41 & 1.01\% & 0.06 & 3.50 & 0.12\% \\
dart & 9.85 & 2210.66 & 0.78\% & 0.04 & 9.85 & 0.09\% \\
jupyter-scripts-dedup-filtered & 7.03 & 910.54 & 0.56\% & 0.03 & 4.06 & 0.06\% \\
jupyter-structured-clean-dedup & 5.72 & 652.90 & 0.45\% & 0.03 & 2.91 & 0.05\% \\
tex & 5.70 & 493.84 & 0.45\% & 0.03 & 2.20 & 0.05\% \\
restructuredtext & 5.43 & 1036.81 & 0.43\% & 0.02 & 4.62 & 0.05\% \\
haskell & 5.41 & 1120.82 & 0.43\% & 0.02 & 4.99 & 0.05\% \\
ada & 0.53 & 56.26 & 0.04\% & 0.02 & 2.35 & 0.05\% \\
yaml & 3.25 & 2295.73 & 0.26\% & 0.01 & 10.23 & 0.03\% \\
common-lisp & 3.25 & 202.43 & 0.26\% & 0.01 & 0.90 & 0.03\% \\
pascal & 3.18 & 202.85 & 0.25\% & 0.01 & 0.90 & 0.03\% \\
makefile & 2.86 & 914.46 & 0.23\% & 0.01 & 4.08 & 0.03\% \\
java-server-pages & 2.79 & 519.49 & 0.22\% & 0.01 & 2.32 & 0.03\% \\
ocaml & 2.27 & 289.99 & 0.18\% & 0.01 & 1.29 & 0.02\% \\
fortran & 2.26 & 184.09 & 0.18\% & 0.01 & 0.82 & 0.02\% \\
visual-basic & 2.24 & 231.06 & 0.18\% & 0.01 & 1.03 & 0.02\% \\
mathematica & 2.10 & 46.81 & 0.17\% & 0.01 & 0.21 & 0.02\% \\
erlang & 1.99 & 253.50 & 0.16\% & 0.01 & 1.13 & 0.02\% \\
vhdl & 1.82 & 124.92 & 0.14\% & 0.01 & 0.56 & 0.02\% \\
elixir & 1.70 & 511.67 & 0.13\% & 0.01 & 2.28 & 0.02\% \\
groovy & 1.69 & 401.75 & 0.13\% & 0.01 & 1.79 & 0.02\% \\
smalltalk & 1.48 & 482.43 & 0.12\% & 0.01 & 2.15 & 0.01\% \\
		\hline
	\end{tabular}
 \label{tab:data_overview_1}
\end{table}

\subsection{Model}

\subsubsection{Tokenizer}

To enhance adaptability to the Chinese programming context, we enriched the StarCoder vocabulary with a substantial addition of Chinese lexicon. Specifically, we collected one million Chinese code files and coding question-and-answer data in Chinese. Utilizing the Tokenizer library from Hugging Face, we identified 40,000 high-frequency Chinese vocabularies. These were merged with the 30,000 high-frequency English vocabularies from the StarCoder lexicon, resulting in the comprehensive CodeShell vocabulary. Experimental results indicate that, compared to the StarCoder vocabulary, CodeShell's lexicon demonstrates a significant advantage in the efficiency of tokenizing Chinese coding questions and answers.

\begin{table}[hbt]
	\centering
    \caption{
    The vocab size and compression rate of StarCoder and CodeShell.
    }
    \begin{tabular}{lccccc}
    \hline
    Tokenizer &	Size & Chinese & English & Code & Total \\
    \hline
StarCoder & 49152 & 1.22 & 3.47 & 3.30 & 2.66 \\
CodeShell & 70020 & 1.50 & 3.47 & 3.30 & 2.95 \\
		\hline
	\end{tabular}
 \label{tab:vocabulary}
\end{table}

\subsubsection{Architecture}

The CodeShell leverages GPT-2~\citep{gpt2} as its foundational architecture, employing advanced techniques such as Grouped-Query Attention~\citep{gqa} and RoPE (Rotary Positional Encoding)~\citep{rope}. The Grouped-Query Attention mechanism optimizes attention operations by clustering similar queries, thereby improving computational efficiency and reducing redundancy. Meanwhile, the Rotary Positional Encoding introduces a more dynamic way of representing the positions of elements in sequences, providing the model with a better grasp of the order and structure within code snippets. 

\begin{table}[hbt]
	\centering
    \caption{
    The Hyperparameter of StarCoder.
    }
    \begin{tabular}{cc}
    \hline
    Hyperparameter & Value \\
    \hline
Hidden size & 4096 \\
Intermediate size & 16384 \\
Hidden layers number & 42 \\
Attention heads number & 32 \\
Query groups number & 8 \\
Max. position embeddings & 8192\\
Position embedding & ROPE\\
Attention & Grouped-query Attention \\
		\hline
	\end{tabular}
 \label{tab:Hyperparameter}
\end{table}

\subsection{Training}

\subsubsection{Optimization}

In our training setup, we opted for the AdamW as our optimizer, configuring it with $\beta 1$ and $\beta 2$ parameters set to 0.9 and 0.95, respectively.
Our training process incorporates a cosine annealing schedule that begins with 1000 warm-up steps, after which the learning rate decreases from 3e-4 to 3e-5 in 127k iterations. 
For our training batches, we process 4 million tokens at one batch, dividing them into sequences of 2048 or 8192 tokens each. 

\subsubsection{Pre-Training}

\begin{figure}[hbt]
\centering
\includegraphics[width=0.8\linewidth]{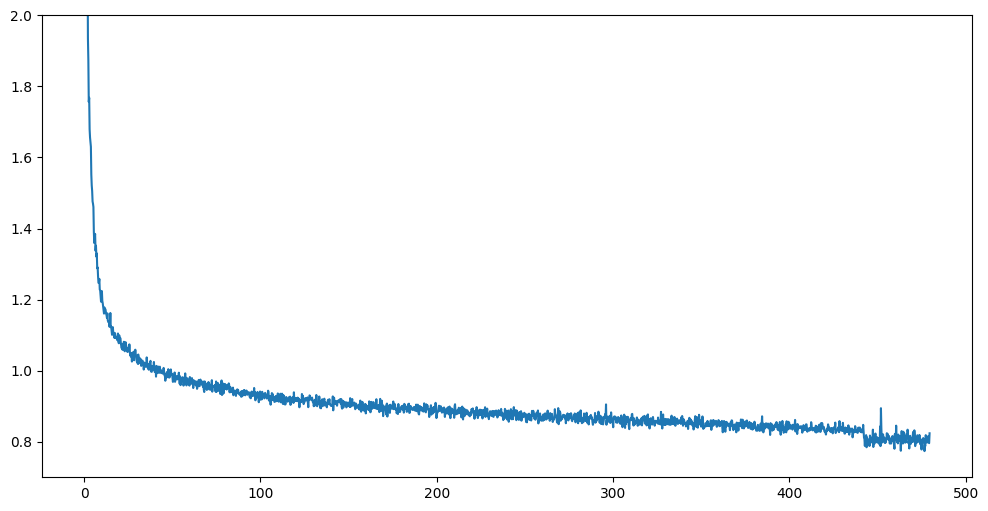}
\caption{Training loss over train tokens.}
\label{fig:losses}
\end{figure}

To balance efficiency and the need for longer context, we initially selected a context length of 2048 for the early stages of pre-training, and after training on nearly 450 billion tokens, we increased the context length to 8192. Following this modification, the GPU throughput decreased from 3200 tokens per GPU per second to 2600 tokens per GPU per second. Concurrently, we conducted experiments to compare the model's performance before and after the change in context length. The results showed that the model's performance remained essentially unchanged, if not improved. The pre-training loss curve is illustrated in the figure \ref{fig:losses}. As illustrated in Figure \ref{fig:losses}, we observed a notable decrease in loss when the context length was expanded from 2048 to 8192. This reduction may stem from the longer context providing more information, thereby simplifying the prediction process. However, it's crucial to highlight that despite the decrease in loss, the model's evaluation metrics did not show any improvement.

\section{Results}

This section explores the performance of CodeShell. CodeShell is evaluated against the latest advanced large language models:

\begin{itemize}

\item \textbf{StarCoder-7B and StarCoder-15B}~\citep{starcoder}, a model with an extensive 7 billion and 15 billion parameter size, is openly available and excels in a variety of programming tasks. It benefits from a carefully selected portion of the Stack dataset, which spans 86 programming languages.

\item \textbf{CodeLlama}~\citep{codellama}, includes a range of code-focused Large Language Models (LLMs), evolving from the LLaMA2~\citep{llama2} models. CodeLlama models are enhanced through ongoing training on a substantial 500 billion token code corpus, leveraging the LLaMA2 base architecture.

\end{itemize}

\subsection{Code generation}

\subsection{Python Code generation}

In this section, we evaluate the performance of CodeShell on Python, comparing it to both open-access and closed-access models. We start by reporting results for Python code generation using the HumanEval~\citep{codex}, MBPP~\citep{mbpp} benchmarks. Results are summarized in Tables \ref{table:humaneval_mbpp}.  The HumanEval dataset is comprised of 164 manually crafted Python tasks, each verified through test cases to evaluate the performance of a Code LLM in generating solutions without prior examples (zero-shot). Conversely, the MBPP benchmark encompasses 500 Python challenges designed to test the model's ability in a context where a small number of examples are provided beforehand (few-shot setting).

The experimental results reveal that CodeShell-7B sets a new benchmark in performance among its peer group, registering an impressive average accuracy of 34.3\% on the HumanEval dataset and 38.7\% on the MBPP benchmark. This performance not only establishes CodeShell-7B as the leader in its class but also underscores its superiority over Code-LLaMA-7B and StarCoder-Base 7B, a similarly scaled open-source model. Specifically, CodeShell-7B shows a robust competitive advantage, even when juxtaposed against larger and more intricate coding models that boast a substantial increase in parameter count.

Please note that, by applying strict criteria for selecting high-quality data and conducting repeated training over multiple epochs, CodeShell has achieved exceptional performance in basic tasks like HumanEval. However, the current data selection and training strategy may fall short for more complex challenges like CoderUJB due to their increased structural and logical complexity. Nonetheless, by refining our data selection process to better align with the demands of such complex tasks, we can still improve the CodeShell's performance on difficult tasks like CoderUJB.

\begin{table}[hbt]
	\centering
    \caption{Performance of approaches on the HumanEval and MBPP.}
	\begin{tabular}{lcc}
		\hline 
		\textbf{Model} &  \textbf{Huamn-eval} & \textbf{MBPP} \\
		\hline
LLaMA-7B & 10.5 & 17.7 \\
LaMDA-137B & 14.0 & 14.8 \\
LLaMA-13B & 15.8 & 22.0 \\
LLaMA-33B & 21.7 & 30.2 \\
CodeGeeX-6B & 22.9 & 24.4 \\
LLaMA-65B & 23.7 & 37.7 \\
PaLM-540B & 26.2 & 36.8 \\
StarCoderBase-7B & 29.4 & 37.6 \\
StarCoderBase-15B & 30.4 & \textbf{49.0} \\
code-cushman-001-12B & 33.5 & 45.9 \\
CodeShell-7B & \textbf{34.3} & 38.7 \\
		\hline
	\end{tabular}
	\label{table:humaneval_mbpp}
\end{table}

\subsection{Multilingual Code generation}

Subsequently, we conduct evaluations of our models across a broader spectrum of programming languages, utilizing the MultiPL\-E~\citep{multiple} benchmark developed by Cassano et al., in 2022 for this purpose. 
The outcomes for a variety of languages, including JavaScript, Java, swift, PHP, et al. are detailed in Table \ref{tb:results_multiple}.

We observed that CodeShell achieved better results in multiple mainstream languages, including JS, Java, and CPP, compared to CodeLlama-7b, StarCoder-7b and StarCoder-15b. However, its performance was inferior in smaller-scale languages like D, Julia, and Lua. We speculate that this is due to the easier accessibility of high-quality corpora for mainstream languages, allowing the model to achieve better results with sufficient data training. Additionally, it is noteworthy that our model demonstrated competitive performance with StarCoder-15b, indicating that pre-training large code models on smaller models holds potential.

\begin{table}[ht]
	\centering
    \caption{
    Performance of approaches on the MultiPL-E.
    }
	\begin{tabular}{lcccc}
		\hline 
		\textbf{Datasets} &  \textbf{CodeShell-7b} & \textbf{CodeLlama-7b} & \textbf{StarCoder-7b} & \textbf{StarCoder-15b} \\
		\hline
            multiple-js	 & \textbf{33.17} & 31.30 & 27.02 & 30.50 \\ 
            multiple-java	 & \textbf{30.43} & 29.24 & 24.30 & 28.16 \\ 
            multiple-cpp	 & 28.21 & 27.33 & 23.04 & \textbf{29.69} \\ 
            multiple-swift & 24.30 & \textbf{25.32} & 15.70 & 18.86 \\ 
            multiple-php	 & \textbf{30.87} & 25.96 & 22.11 & 26.65 \\ 
            multiple-d	 & 8.85 & 11.60 & 8.08 & \textbf{13.33} \\ 
            multiple-jl	 & 22.08 & \textbf{25.28} & 22.96 & 22.20 \\ 
            multiple-lua	 & 22.39 & \textbf{30.50} & 22.92 & 23.35 \\ 
            multiple-r	 & \textbf{20.52} & 18.57 & 14.29 & 15.90 \\ 
            multiple-rkt	 & \textbf{17.20} & 12.55 & 10.43 & 9.94 \\ 
            multiple-rs	 & 24.55 & \textbf{25.90} & 22.82 & 21.92 \\ 
		\hline
	\end{tabular}
	\label{tb:results_multiple}
\end{table}

\subsection{Code Completion}

During the pre-training phase, CodeShell-7B models were trained with a 0.5 Fill-In-the-Middle~\citep{fim} (FIM) rate, a technique that enhances their ability to generate code by filling in gaps based on the context surrounding a code snippet. This approach proves particularly useful for code completion applications. Similar capabilities are seen in open-source models such as StarCoder-15B and CodeLlama-7B. Followed by Allal et al.~\citep{santacoder}, , we mask out a single line of text from the function body, and prompt the model to fill in that line of code. We evaluate on the MultiPL\-E~\citep{multiple} benchmark across the three programming languages, and use the single-line exact match ~\citep{incoder} as the metric.

The results, meticulously detailed in Table \ref{tab:results_code_completion}, illustrate that CodeShell-7B surpasses both StarCoder and CodeLlama in these benchmarks, highlighting the critical role of high-quality pre-training data. Similar to CodeLlama, CodeShell employs a context length of 2048 during the initial stages of pre-training to enhance training efficiency. In the final stages of training, specifically the last 50 billion tokens, CodeShell extends the context length from 2048 to 8192. The experimental findings reveal that the capability to handle longer code sequences can be achieved with only a minimal amount of training data. Hence, during the pre-training process, we can initially utilize a smaller context length to boost efficiency without compromising the model's eventual proficiency in managing extensive code.

\begin{table}[ht]
	\centering
    \caption{
    Performance of approaches on the Code Completion.
    }
	\begin{tabular}{lcccc}
		\hline 
		\textbf{Model} &  \textbf{python} & \textbf{java} & \textbf{javascript} & \textbf{Mean} \\
		\hline
            StarCoder-15B & 62.0 & 73.0 & 74.0 & 69.7 \\ 
            CodeLlama-7B  & \textbf{67.6} & 74.3 & 80.2 & 74.0 \\ 
            CodeShell-7B  & 65.3 & \textbf{76.9} & \textbf{80.6} & \textbf{74.3} \\
		\hline
	\end{tabular}
	\label{tab:results_code_completion}
\end{table}




\subsection{Data Ablations}

To validate the effectiveness of our high-quality code data filtering mechanism, we constructed two datasets: a random dataset, which was created by randomly sampling 2 billion tokens from the deduplicated original dataset, and a filtered dataset, which was compiled by sampling 2 billion tokens from high to low based on scores from a high-quality scorer. 
Specifically, to expedite the validation of different architectures on model performance, we developed a "codeshell-small" version with 24 layers, a hidden size of 2048, and a total of 1 billion parameters.
We conducted experiments on these two datasets using the codeshell-small model, and the results are depicted in the figure below. The findings demonstrate that the model trained on the filtered dataset consistently outperformed the model trained on the random dataset in terms of performance. Moreover, the final performance of the filtered model showed nearly a 100\% improvement over the random model. These results further underscore the critical role of data quality in the training of large models and also validate the effectiveness of our data filtering approach.

\begin{figure}[hbt]
\centering
\includegraphics[width=0.6\linewidth]{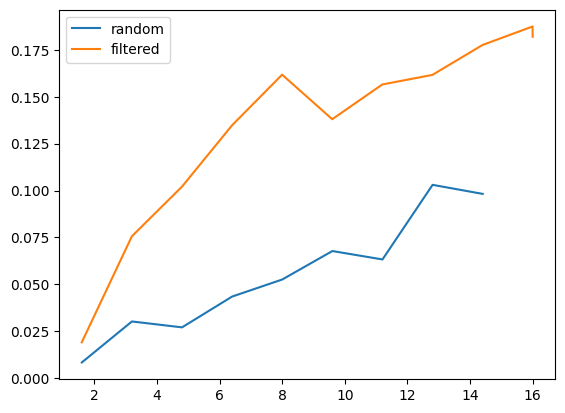}
\caption{The effectiveness of high-quality code data filtering mechanism.}
\end{figure}

\section{Conclusion}

We have introduced CodeShell, a large-scale code model. In this paper, we explored the impact of different model architectures and data filtering strategies on model performance. The experimental results indicate that high-quality data remains the most significant factor affecting the performance of large models. Through the selection of high-quality data, CodeShell achieved optimal performance across a variety of programming languages at a comparable scale. Moreover, in experiments with high-quality data, the data filtering method we proposed achieved a 100\% performance improvement over random data selection.

\newpage

\bibliographystyle{iclr2023_conference}
\bibliography{main}

\begin{thebibliography}{22}
\providecommand{\natexlab}[1]{#1}
\providecommand{\url}[1]{\texttt{#1}}
\expandafter\ifx\csname urlstyle\endcsname\relax
  \providecommand{\doi}[1]{doi: #1}\else
  \providecommand{\doi}{doi: \begingroup \urlstyle{rm}\Url}\fi

\bibitem[Ainslie et~al.(2023)Ainslie, Lee-Thorp, de~Jong, Zemlyanskiy, Lebr{\'o}n, and Sanghai]{gqa}
Joshua Ainslie, James Lee-Thorp, Michiel de~Jong, Yury Zemlyanskiy, Federico Lebr{\'o}n, and Sumit Sanghai.
\newblock Gqa: Training generalized multi-query transformer models from multi-head checkpoints.
\newblock \emph{arXiv preprint arXiv:2305.13245}, 2023.

\bibitem[Allal et~al.(2023)Allal, Li, Kocetkov, Mou, Akiki, Ferrandis, Muennighoff, Mishra, Gu, Dey, et~al.]{santacoder}
Loubna~Ben Allal, Raymond Li, Denis Kocetkov, Chenghao Mou, Christopher Akiki, Carlos~Munoz Ferrandis, Niklas Muennighoff, Mayank Mishra, Alex Gu, Manan Dey, et~al.
\newblock Santacoder: don't reach for the stars!
\newblock \emph{arXiv preprint arXiv:2301.03988}, 2023.

\bibitem[Austin et~al.(2021)Austin, Odena, Nye, Bosma, Michalewski, Dohan, Jiang, Cai, Terry, Le, and Sutton]{mbpp}
Jacob Austin, Augustus Odena, Maxwell~I. Nye, Maarten Bosma, Henryk Michalewski, David Dohan, Ellen Jiang, Carrie~J. Cai, Michael Terry, Quoc~V. Le, and Charles Sutton.
\newblock Program synthesis with large language models.
\newblock \emph{CoRR}, abs/2108.07732, 2021.
\newblock URL \url{https://arxiv.org/abs/2108.07732}.

\bibitem[Bavarian et~al.(2022)Bavarian, Jun, Tezak, Schulman, McLeavey, Tworek, and Chen]{fim}
Mohammad Bavarian, Heewoo Jun, Nikolas Tezak, John Schulman, Christine McLeavey, Jerry Tworek, and Mark Chen.
\newblock Efficient training of language models to fill in the middle.
\newblock \emph{arXiv preprint arXiv:2207.14255}, 2022.

\bibitem[Cassano et~al.(2023)Cassano, Gouwar, Nguyen, Nguyen, Phipps-Costin, Pinckney, Yee, Zi, Anderson, Feldman, et~al.]{multiple}
Federico Cassano, John Gouwar, Daniel Nguyen, Sydney Nguyen, Luna Phipps-Costin, Donald Pinckney, Ming-Ho Yee, Yangtian Zi, Carolyn~Jane Anderson, Molly~Q Feldman, et~al.
\newblock Multipl-e: a scalable and polyglot approach to benchmarking neural code generation.
\newblock \emph{IEEE Transactions on Software Engineering}, 2023.

\bibitem[Chen et~al.(2021)Chen, Tworek, Jun, Yuan, de~Oliveira~Pinto, Kaplan, Edwards, Burda, Joseph, Brockman, Ray, Puri, Krueger, Petrov, Khlaaf, Sastry, Mishkin, Chan, Gray, Ryder, Pavlov, Power, Kaiser, Bavarian, Winter, Tillet, Such, Cummings, Plappert, Chantzis, Barnes, Herbert{-}Voss, Guss, Nichol, Paino, Tezak, Tang, Babuschkin, Balaji, Jain, Saunders, Hesse, Carr, Leike, Achiam, Misra, Morikawa, Radford, Knight, Brundage, Murati, Mayer, Welinder, McGrew, Amodei, McCandlish, Sutskever, and Zaremba]{codex}
Mark Chen, Jerry Tworek, Heewoo Jun, Qiming Yuan, Henrique~Pond{\'{e}} de~Oliveira~Pinto, Jared Kaplan, Harrison Edwards, Yuri Burda, Nicholas Joseph, Greg Brockman, Alex Ray, Raul Puri, Gretchen Krueger, Michael Petrov, Heidy Khlaaf, Girish Sastry, Pamela Mishkin, Brooke Chan, Scott Gray, Nick Ryder, Mikhail Pavlov, Alethea Power, Lukasz Kaiser, Mohammad Bavarian, Clemens Winter, Philippe Tillet, Felipe~Petroski Such, Dave Cummings, Matthias Plappert, Fotios Chantzis, Elizabeth Barnes, Ariel Herbert{-}Voss, William~Hebgen Guss, Alex Nichol, Alex Paino, Nikolas Tezak, Jie Tang, Igor Babuschkin, Suchir Balaji, Shantanu Jain, William Saunders, Christopher Hesse, Andrew~N. Carr, Jan Leike, Joshua Achiam, Vedant Misra, Evan Morikawa, Alec Radford, Matthew Knight, Miles Brundage, Mira Murati, Katie Mayer, Peter Welinder, Bob McGrew, Dario Amodei, Sam McCandlish, Ilya Sutskever, and Wojciech Zaremba.
\newblock Evaluating large language models trained on code.
\newblock \emph{CoRR}, abs/2107.03374, 2021.
\newblock URL \url{https://arxiv.org/abs/2107.03374}.

\bibitem[Christiani \& Pagh(2017)Christiani and Pagh]{minihash}
Tobias Christiani and Rasmus Pagh.
\newblock Set similarity search beyond minhash.
\newblock In \emph{Proceedings of the 49th annual ACM SIGACT symposium on theory of computing}, pp.\  1094--1107, 2017.

\bibitem[Fried et~al.(2023)Fried, Aghajanyan, Lin, Wang, Wallace, Shi, Zhong, Yih, Zettlemoyer, and Lewis]{incoder}
Daniel Fried, Armen Aghajanyan, Jessy Lin, Sida Wang, Eric Wallace, Freda Shi, Ruiqi Zhong, Scott Yih, Luke Zettlemoyer, and Mike Lewis.
\newblock Incoder: {A} generative model for code infilling and synthesis.
\newblock In \emph{The Eleventh International Conference on Learning Representations, {ICLR} 2023, Kigali, Rwanda, May 1-5, 2023}. OpenReview.net, 2023.
\newblock URL \url{https://openreview.net/pdf?id=hQwb-lbM6EL}.

\bibitem[{Github Archive}(2024)]{gharchive}
{Github Archive}, 2024.
\newblock URL \url{https://gharchive.org}.

\bibitem[Gunasekar et~al.(2023)Gunasekar, Zhang, Aneja, Mendes, Del~Giorno, Gopi, Javaheripi, Kauffmann, de~Rosa, Saarikivi, et~al.]{phi1}
Suriya Gunasekar, Yi~Zhang, Jyoti Aneja, Caio C{\'e}sar~Teodoro Mendes, Allie Del~Giorno, Sivakanth Gopi, Mojan Javaheripi, Piero Kauffmann, Gustavo de~Rosa, Olli Saarikivi, et~al.
\newblock Textbooks are all you need.
\newblock \emph{arXiv preprint arXiv:2306.11644}, 2023.

\bibitem[Jansen et~al.(2022)Jansen, Tong, Zevallos, and Suarez]{pplf}
Tim Jansen, Yangling Tong, Victoria Zevallos, and Pedro~Ortiz Suarez.
\newblock Perplexed by quality: A perplexity-based method for adult and harmful content detection in multilingual heterogeneous web data.
\newblock \emph{arXiv preprint arXiv:2212.10440}, 2022.

\bibitem[Just et~al.(2014)Just, Jalali, and Ernst]{defects4j}
Ren{\'{e}} Just, Darioush Jalali, and Michael~D. Ernst.
\newblock Defects4j: a database of existing faults to enable controlled testing studies for java programs.
\newblock In Corina~S. Pasareanu and Darko Marinov (eds.), \emph{International Symposium on Software Testing and Analysis, {ISSTA} '14, San Jose, CA, {USA} - July 21 - 26, 2014}, pp.\  437--440. {ACM}, 2014.
\newblock \doi{10.1145/2610384.2628055}.

\bibitem[Kocetkov et~al.(2022)Kocetkov, Li, Allal, Li, Mou, Ferrandis, Jernite, Mitchell, Hughes, Wolf, Bahdanau, von Werra, and de~Vries]{thestack}
Denis Kocetkov, Raymond Li, Loubna~Ben Allal, Jia Li, Chenghao Mou, Carlos~Mu{\~{n}}oz Ferrandis, Yacine Jernite, Margaret Mitchell, Sean Hughes, Thomas Wolf, Dzmitry Bahdanau, Leandro von Werra, and Harm de~Vries.
\newblock The stack: 3 {TB} of permissively licensed source code.
\newblock \emph{CoRR}, abs/2211.15533, 2022.
\newblock \doi{10.48550/ARXIV.2211.15533}.
\newblock URL \url{https://doi.org/10.48550/arXiv.2211.15533}.

\bibitem[Li et~al.(2023)Li, Allal, Zi, Muennighoff, Kocetkov, Mou, Marone, Akiki, Li, Chim, Liu, Zheltonozhskii, Zhuo, Wang, Dehaene, Davaadorj, Lamy{-}Poirier, Monteiro, Shliazhko, Gontier, Meade, Zebaze, Yee, Umapathi, Zhu, Lipkin, Oblokulov, Wang, V, Stillerman, Patel, Abulkhanov, Zocca, Dey, Zhang, Moustafa{-}Fahmy, Bhattacharyya, Yu, Singh, Luccioni, Villegas, Kunakov, Zhdanov, Romero, Lee, Timor, Ding, Schlesinger, Schoelkopf, Ebert, Dao, Mishra, Gu, Robinson, Anderson, Dolan{-}Gavitt, Contractor, Reddy, Fried, Bahdanau, Jernite, Ferrandis, Hughes, Wolf, Guha, von Werra, and de~Vries]{starcoder}
Raymond Li, Loubna~Ben Allal, Yangtian Zi, Niklas Muennighoff, Denis Kocetkov, Chenghao Mou, Marc Marone, Christopher Akiki, Jia Li, Jenny Chim, Qian Liu, Evgenii Zheltonozhskii, Terry~Yue Zhuo, Thomas Wang, Olivier Dehaene, Mishig Davaadorj, Joel Lamy{-}Poirier, Jo{\~{a}}o Monteiro, Oleh Shliazhko, Nicolas Gontier, Nicholas Meade, Armel Zebaze, Ming{-}Ho Yee, Logesh~Kumar Umapathi, Jian Zhu, Benjamin Lipkin, Muhtasham Oblokulov, Zhiruo Wang, Rudra~Murthy V, Jason Stillerman, Siva~Sankalp Patel, Dmitry Abulkhanov, Marco Zocca, Manan Dey, Zhihan Zhang, Nour Moustafa{-}Fahmy, Urvashi Bhattacharyya, Wenhao Yu, Swayam Singh, Sasha Luccioni, Paulo Villegas, Maxim Kunakov, Fedor Zhdanov, Manuel Romero, Tony Lee, Nadav Timor, Jennifer Ding, Claire Schlesinger, Hailey Schoelkopf, Jan Ebert, Tri Dao, Mayank Mishra, Alex Gu, Jennifer Robinson, Carolyn~Jane Anderson, Brendan Dolan{-}Gavitt, Danish Contractor, Siva Reddy, Daniel Fried, Dzmitry Bahdanau, Yacine Jernite, Carlos~Mu{\~{n}}oz Ferrandis, Sean Hughes, Thomas
  Wolf, Arjun Guha, Leandro von Werra, and Harm de~Vries.
\newblock Starcoder: may the source be with you!
\newblock \emph{CoRR}, abs/2305.06161, 2023.
\newblock \doi{10.48550/ARXIV.2305.06161}.
\newblock URL \url{https://doi.org/10.48550/arXiv.2305.06161}.

\bibitem[Luo et~al.(2023)Luo, Xu, Zhao, Sun, Geng, Hu, Tao, Ma, Lin, and Jiang]{wizardcoder}
Ziyang Luo, Can Xu, Pu~Zhao, Qingfeng Sun, Xiubo Geng, Wenxiang Hu, Chongyang Tao, Jing Ma, Qingwei Lin, and Daxin Jiang.
\newblock Wizardcoder: Empowering code large language models with evol-instruct.
\newblock \emph{arXiv preprint arXiv:2306.08568}, 2023.

\bibitem[Nijkamp et~al.(2023)Nijkamp, Pang, Hayashi, Tu, Wang, Zhou, Savarese, and Xiong]{codegen}
Erik Nijkamp, Bo~Pang, Hiroaki Hayashi, Lifu Tu, Huan Wang, Yingbo Zhou, Silvio Savarese, and Caiming Xiong.
\newblock Codegen: An open large language model for code with multi-turn program synthesis.
\newblock In \emph{The Eleventh International Conference on Learning Representations, {ICLR} 2023, Kigali, Rwanda, May 1-5, 2023}. OpenReview.net, 2023.
\newblock URL \url{https://openreview.net/pdf?id=iaYcJKpY2B\_}.

\bibitem[OpenAI(2023)]{gpt4report}
OpenAI.
\newblock {GPT-4} technical report.
\newblock \emph{CoRR}, abs/2303.08774, 2023.
\newblock \doi{10.48550/ARXIV.2303.08774}.
\newblock URL \url{https://doi.org/10.48550/arXiv.2303.08774}.

\bibitem[Radford et~al.(2019)Radford, Wu, Child, Luan, Amodei, Sutskever, et~al.]{gpt2}
Alec Radford, Jeffrey Wu, Rewon Child, David Luan, Dario Amodei, Ilya Sutskever, et~al.
\newblock Language models are unsupervised multitask learners.
\newblock \emph{OpenAI blog}, 1\penalty0 (8):\penalty0 9, 2019.

\bibitem[Rozi{\`{e}}re et~al.(2023)Rozi{\`{e}}re, Gehring, Gloeckle, Sootla, Gat, Tan, Adi, Liu, Remez, Rapin, Kozhevnikov, Evtimov, Bitton, Bhatt, Canton{-}Ferrer, Grattafiori, Xiong, D{\'{e}}fossez, Copet, Azhar, Touvron, Martin, Usunier, Scialom, and Synnaeve]{codellama}
Baptiste Rozi{\`{e}}re, Jonas Gehring, Fabian Gloeckle, Sten Sootla, Itai Gat, Xiaoqing~Ellen Tan, Yossi Adi, Jingyu Liu, Tal Remez, J{\'{e}}r{\'{e}}my Rapin, Artyom Kozhevnikov, Ivan Evtimov, Joanna Bitton, Manish Bhatt, Cristian Canton{-}Ferrer, Aaron Grattafiori, Wenhan Xiong, Alexandre D{\'{e}}fossez, Jade Copet, Faisal Azhar, Hugo Touvron, Louis Martin, Nicolas Usunier, Thomas Scialom, and Gabriel Synnaeve.
\newblock Code llama: Open foundation models for code.
\newblock \emph{CoRR}, abs/2308.12950, 2023.
\newblock \doi{10.48550/ARXIV.2308.12950}.
\newblock URL \url{https://doi.org/10.48550/arXiv.2308.12950}.

\bibitem[Su et~al.(2024)Su, Ahmed, Lu, Pan, Bo, and Liu]{rope}
Jianlin Su, Murtadha Ahmed, Yu~Lu, Shengfeng Pan, Wen Bo, and Yunfeng Liu.
\newblock Roformer: Enhanced transformer with rotary position embedding.
\newblock \emph{Neurocomputing}, 568:\penalty0 127063, 2024.

\bibitem[Touvron et~al.(2023)Touvron, Martin, Stone, Albert, Almahairi, Babaei, Bashlykov, Batra, Bhargava, Bhosale, Bikel, Blecher, Canton{-}Ferrer, Chen, Cucurull, Esiobu, Fernandes, Fu, Fu, Fuller, Gao, Goswami, Goyal, Hartshorn, Hosseini, Hou, Inan, Kardas, Kerkez, Khabsa, Kloumann, Korenev, Koura, Lachaux, Lavril, Lee, Liskovich, Lu, Mao, Martinet, Mihaylov, Mishra, Molybog, Nie, Poulton, Reizenstein, Rungta, Saladi, Schelten, Silva, Smith, Subramanian, Tan, Tang, Taylor, Williams, Kuan, Xu, Yan, Zarov, Zhang, Fan, Kambadur, Narang, Rodriguez, Stojnic, Edunov, and Scialom]{llama2}
Hugo Touvron, Louis Martin, Kevin Stone, Peter Albert, Amjad Almahairi, Yasmine Babaei, Nikolay Bashlykov, Soumya Batra, Prajjwal Bhargava, Shruti Bhosale, Dan Bikel, Lukas Blecher, Cristian Canton{-}Ferrer, Moya Chen, Guillem Cucurull, David Esiobu, Jude Fernandes, Jeremy Fu, Wenyin Fu, Brian Fuller, Cynthia Gao, Vedanuj Goswami, Naman Goyal, Anthony Hartshorn, Saghar Hosseini, Rui Hou, Hakan Inan, Marcin Kardas, Viktor Kerkez, Madian Khabsa, Isabel Kloumann, Artem Korenev, Punit~Singh Koura, Marie{-}Anne Lachaux, Thibaut Lavril, Jenya Lee, Diana Liskovich, Yinghai Lu, Yuning Mao, Xavier Martinet, Todor Mihaylov, Pushkar Mishra, Igor Molybog, Yixin Nie, Andrew Poulton, Jeremy Reizenstein, Rashi Rungta, Kalyan Saladi, Alan Schelten, Ruan Silva, Eric~Michael Smith, Ranjan Subramanian, Xiaoqing~Ellen Tan, Binh Tang, Ross Taylor, Adina Williams, Jian~Xiang Kuan, Puxin Xu, Zheng Yan, Iliyan Zarov, Yuchen Zhang, Angela Fan, Melanie Kambadur, Sharan Narang, Aur{\'{e}}lien Rodriguez, Robert Stojnic, Sergey Edunov,
  and Thomas Scialom.
\newblock Llama 2: Open foundation and fine-tuned chat models.
\newblock \emph{CoRR}, abs/2307.09288, 2023.
\newblock \doi{10.48550/ARXIV.2307.09288}.
\newblock URL \url{https://doi.org/10.48550/arXiv.2307.09288}.

\bibitem[Yu et~al.(2023)Yu, Shen, Ran, Zhang, Zhang, Ma, Liang, Li, Xie, and Wang]{codereval}
Hao Yu, Bo~Shen, Dezhi Ran, Jiaxin Zhang, Qi~Zhang, Yuchi Ma, Guangtai Liang, Ying Li, Tao Xie, and Qianxiang Wang.
\newblock Codereval: {A} benchmark of pragmatic code generation with generative pre-trained models.
\newblock \emph{CoRR}, abs/2302.00288, 2023.
\newblock \doi{10.48550/ARXIV.2302.00288}.
\newblock URL \url{https://doi.org/10.48550/arXiv.2302.00288}.

\end{thebibliography}

\end{document}